\documentclass[english,twocolumn,prl]{revtex4-2}
\usepackage[latin9]{inputenc}
\usepackage{xcolor}
\usepackage{babel}
\usepackage{amsmath}
\usepackage{amssymb}
\usepackage{graphicx}
\usepackage{lipsum}  
\usepackage{mathtools}
\usepackage{url}
\usepackage{comment}
\usepackage{bbold}

\usepackage{hyperref}

\global\long\def\ket#1{\left| #1\right\rangle }%
\global\long\def\bra#1{\left\langle #1 \right|}%
\global\long\def\av#1{\left\langle #1 \right\rangle }%
\global\long\def\tr{\text{tr}}%
\global\long\def\im{\text{Im}}%
\global\long\def\dag{\dagger}%
\global\long\def\g{\gamma}%
\global\long\def\G{\Gamma}%
\global\long\def\d{\mathbf{\delta}}%
\global\long\def\S{\Sigma}%
\global\long\def\w{\omega}%

\newcommand{\unige}[0]{$^1$Department of Quantum Matter Physics, \'Ecole de Physique University of Geneva}
\newcommand{\cea}[0]{$^4$IRIG-MEM-L\_Sim, Univ. Grenoble Alpes, CEA, Grenoble INP,  Grenoble 38000, France.}

\makeatletter
\usepackage{amsmath}

\newcommand{\ubcolor}[2]{\textcolor{lightgray}{\underbracket[0.140ex]{\textcolor{black}{#1}}_{\textcolor{black}{#2}}}}

\makeatother

\begin{document}
\title{Exact description of transport and non-reciprocity in monitored quantum devices}
\author{Jo\~ao Ferreira,$^1$ Tony Jin,$^{1,2}$ Jochen Mannhart,$^3$ Thierry Giamarchi,$^1$ and Michele Filippone$^4$}
\affiliation{\unige}
\affiliation{$^2$Pritzker School of Molecular Engineering, University of Chicago, Chicago, Illinois 60637, USA}
\affiliation{$^3$Max Planck Institute for Solid State Research, Heisenbergstr. 1, 70569 Stuttgart,
Germany}
\affiliation{\cea}

\begin{abstract}
We study non-interacting fermionic systems undergoing continuous monitoring and driven by biased reservoirs. Averaging over the measurement outcomes, we derive exact formulas for the particle and heat flows in the system. We show that these currents feature competing elastic and inelastic components, which depend non-trivially on the monitoring strength $\gamma$. We highlight that monitor-induced inelastic processes lead to non-reciprocal currents, allowing to extract work from measurements without active feedback control. We illustrate our formalism with two distinct monitoring schemes providing measurement-induced power or cooling.~Optimal performances are found for values of the monitoring strength $\gamma$ which are hard to address with perturbative approaches.
\end{abstract}

\maketitle


{\it Introduction --} 
Non-unitary dynamics in quantum systems stems from interactions with the environment~\cite{feynman2000theory,CALDEIRA1983587,gardiner2004quantum,RevModPhys.82.1155}, 
which 
 induce dissipation and usually suppress quantum
coherence~\cite{walls_1985,zurech_2003}.~Nonetheless, non-unitary evolution caused by engineered dissipation~\cite{poyatos_reservoir_engineering_1996,Mueller_2012,grimm2020stabilization,lescanne2020exponential,PhysRevLett.109.045302} or measurements~\cite{gottesman1997stabilizer,calderbank_stabilizer_1998} can stabilize target quantum states, 
 many-body correlations~\cite{barontini_2013,luschen_2017,tomita2017observation,fitzpatrick_2017,ma2019dissipatively,yamamoto_2021,dogra2019dissipation,PhysRevX.11.041046,syassen2008strong,Sponselee_2019} and
 exotic entanglement dynamics~\cite{skinner_mesaurement_2019,li_2018,choi_2020,gullans_2020,alberton_entanglement_2021,Mueller_2021,hoke2023quantum,poboiko2023theory}.
 
Of particular interest are the effects of non-unitarity on quantum transport. Environment-assisted processes can drive currents in coherent systems~\cite{Plenio_2008,Rebentrost_2009,caruso_2010,viciani2015observation,maier2019environment,corman_dissipativeQPC_2019,Bredol_2021_alone,Bredol_2021,huang2023superfluid,Visuri_2022_sym,PhysRevB.70.075301} and the impact of losses~\cite{PhysRevA.103.L060201,PhysRevLett.124.147203,PhysRevA.104.053305,mazza2023dissipative,Visuri_2022_sym,Visuri_2022,visuri2023dc,jin_generic_2020,PhysRevLett.122.040402,PhysRevB.104.155431} is investigated in quantum simulators~\cite{syassen2008strong,Sponselee_2019,corman_dissipativeQPC_2019,huang2023superfluid}.  Work extraction from dissipative environments~\cite{RevModPhys.78.217,Benenti_2017} or active monitoring~\cite{Elouard_2017,Elouard_Efficient_2018,naghiloo_information_2018,bresque_two_qubit_2021,stevens_energetics_2022,linpeng_energetic_2022,liu2023maxwell,Annby_Andersson_2022} may use quantum effects at the nanoscale to break the operational limits imposed by classical thermodynamics~\cite{Ryu_2022}.

Quantum devices are usually driven by thermodynamic baths, whose large number of degrees of freedom  challenges exact numerical~\cite{daley_quantum_trajectories_2014} and analytical~\cite{Prosen_2008,znidaric_2010,Sieberer_2016,medvedyeva_exact_2016} approaches, especially to capture the long-time or stationary dynamics of monitored or open settings. Local master equation approaches, based on weak coupling assumptions, may miss interesting effects~\cite{strasberg_2018} or imply apparent violations of the second law of thermodynamics~\cite{whitney2023illusory,Levy_2014,Hofer_2017,Novotny_2002}. 

In this work, we derive exact formulas 
for the particle and heat currents driven by continuous monitoring of a single-particle observable $\mathcal O$  and biased reservoirs in  free fermion systems. We  exploit an exact self-consistent
Born scheme for 2-point correlation functions~\cite{Dolgirev_2020,Jin_2022} and rely on a generalized Meir-Wingreen's approach~\cite{Meir_1992,jin_generic_2020} to account for biased  reservoirs. 
Our main result is formula~\eqref{main}, which offers a simple  and exact tool to address novel quantum transport phenomena in coherent systems under continuous monitoring.

\begin{figure*}[ht!]
\centering{}
\includegraphics[width=\textwidth]{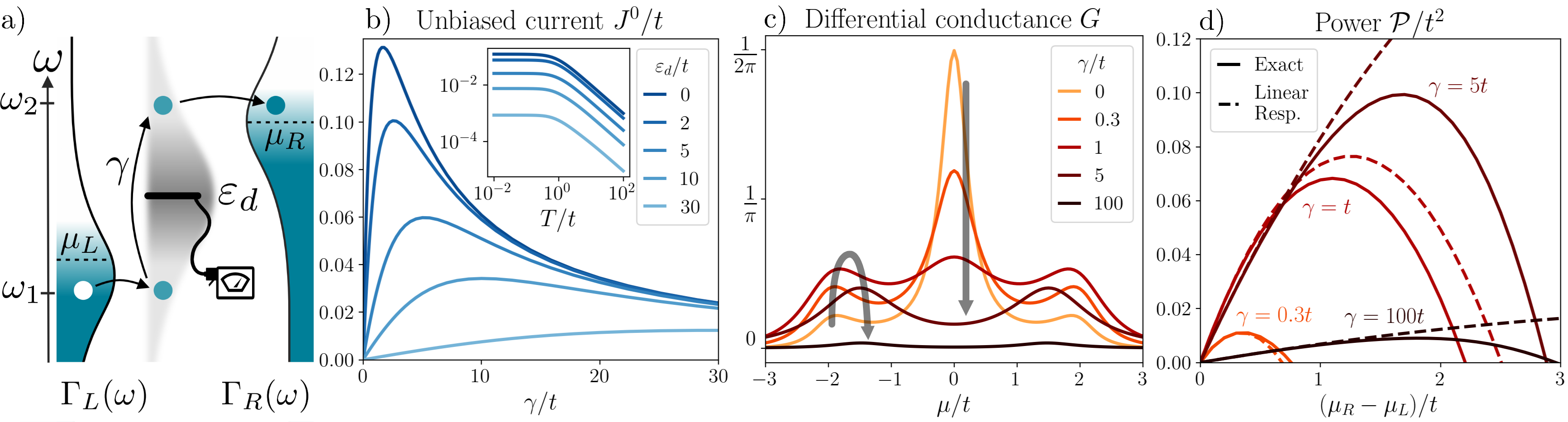}
\caption{a) \label{fig:setup} Monitored level of energy $\varepsilon_d$, coupled to left and right reservoirs
with asymmetric hybridization functions 
$\protect\G_{L}(\protect\w)\neq \protect\G_{R}(\protect\w)$. The level occupation is measured with strength $\gamma$, providing the inelastic mechanism promoting particles from energy $\omega_1$ to $\omega_2$ and inducing a current against the bias (arrows). The blue-shaded areas correspond to the finite-temperature Fermi distributions of the reservoirs. For all plots, we use the two-filter model discussed in the main text, 
with $\varepsilon_R=1.48t=-\varepsilon_L,\,\Delta=0.55t$,  which we found to maximize the unbiased particle current at $\mu_{L,R}=T_{L,R}=0$.
b) \label{fig:anomalous}Peaked structure of the unbiased particle current as a function of the measurement strength $\g$ for varying $\varepsilon_d$. 
Inset: The unbiased current decays monotonously for increasing temperatures ($\g=1$).
c) \label{fig:conductance}Differential conductance  $G$ as a function of the chemical potential $\mu$ at $T=0$ for increasing $\gamma$. The measurement suppresses the resonance associated to the single level and favors those from the filters, as highlighted by arrows. 
d) \label{fig:power} Electric power as function of a symmetric bias $\mu_R-\mu_L$ around $\mu=0$,  for different values of $\g$. Dashed lines correspond to linear response calculations. }
\end{figure*}

\begin{figure*}[ht!]
\centering{}
\includegraphics[width=\textwidth]{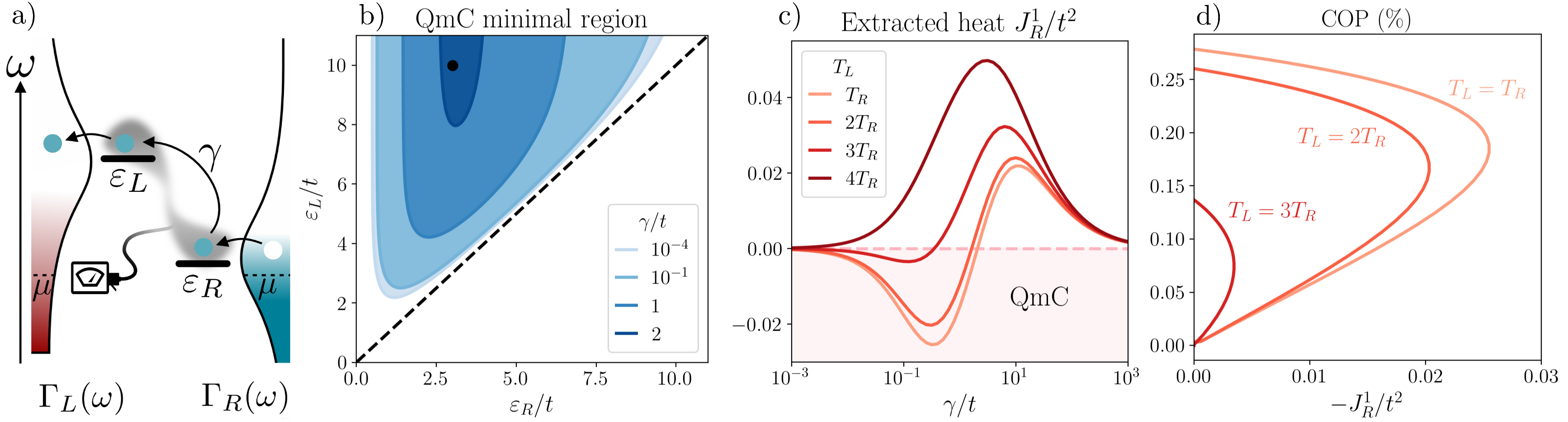}
\caption{
a) \label{fig:model_qssep}  Two-level system under continuous monitoring of its cross-correlations, coupled to a left (hot) and right (cold) reservoir.
For applications, we consider the same filters as in Fig.~\ref{fig:setup}, aligned with the levels  $\varepsilon_{L/R}$.  
b) \label{fig:cbh_region}Parameter region 
where a reservoir at a temperature $T_R=t$  can be cooled by measurement for different values of $\g$ and $\Delta=0.5t$. The range of parameters for which quantum measurement cooling is possible reduces by increasing $\gamma$. The black dot corresponds to $\varepsilon_L=10t$ and $\varepsilon_R=3t$, where panels (c) and (d) are derived. c) \label{fig:heat_qssep} Heat flowing into the right reservoirs for increasing temperature bias $T_L>T_R$. Quantum measurement cooling (QmC) occurs in the colored region and below some critical temperature bias. 
d) \label{fig:cop}Parametric plot of the coefficient of performance (COP) of QmC. Curves are obtained by varying the measurement strength $\g$. }
\end{figure*}

We provide two illustrations of our approach showing monitor-assisted non-reciprocal effects in quantum systems. We consider first the  continuous monitoring  of a single level (Fig.~\ref{fig:setup}). Under generic assumptions, we find that monitoring triggers a non-reciprocal current between reservoirs without external bias, and thus generates power. We then show that monitoring cross-correlations between two sites (Fig.~\ref{fig:model_qssep}) enables quantum measurement cooling~\cite{buffoni_QMC_2019}. 
For both cases, we highlight non-trivial dependencies on the measurement strength $\gamma$, showcased by peaks of performances in regimes which  are not encompassed by perturbative approaches. We also stress that the measurement-based engines described here do not rely on  feedback-loops or Maxwell's  demons~\cite{Elouard_2017,Elouard_Efficient_2018,naghiloo_information_2018,bresque_two_qubit_2021,stevens_energetics_2022,linpeng_energetic_2022,liu2023maxwell,Annby_Andersson_2022}. 


{\it Derivation of monitored currents -- } 
For simplicity, we consider 2-terminal setups~\footnote{Generalizations with more terminals and internal degrees of freedom ({\it e.g.} spin) are straightforward.} described by Hamiltonians of form $\mathcal H=\mathcal H_{\rm res}+\mathcal H_{\rm T}+\mathcal H_{\rm sys}$. Left and right ($r=L/R$) reservoirs are ruled by $\mathcal H_{\rm res}=\sum_{r,k}\varepsilon^{\phantom\dagger}_{r,k}c^\dagger_{r,k}c^{\phantom \dagger}_{r,k}$, where $c_{k,r}$ annihilates fermions of the reservoir $r$ in mode $k$ of  energy $\varepsilon_{r,k}$. 
Both reservoirs are in thermal equilibrium, with chemical potential $\mu_r$, temperature $T_r$, and mode occupation obeying Fermi's distribution $f_r(\varepsilon)=[e^{(\varepsilon-\mu_r)/T_r}+1]^{-1}$.  Free fermions in the system are described by $\mathcal H_{\rm sys}=\sum_{i,j}d^\dagger_i h^{\phantom\dagger}_{ij}d^{\phantom\dagger}_j\,$, where $h_{ij}$ is a single-particle Hamiltonian with  labels $i,j$ referring to internal degrees of freedom (orbitals, spin\ldots). The coupling between system and reservoirs reads $\mathcal H_{\rm T}=\sum_{r,k,i}t^{\phantom\dagger}_{r,ki}c^\dagger_{r,k}d^{\phantom \dagger}_i+\mbox{H.c.}$, where $t_{r,ki}$ are tunnel amplitudes.

When an observable of the system $\mathcal O$ is continuously monitored with strength $\g$, the averaged dynamics of the system density matrix $\rho$ obeys  Lindblad's equation $\partial_t\rho=-i[\mathcal H,\rho]+\mathcal D[\rho]$, where ($\hbar=e=k_{\rm B}=1$)~\cite{Dalibard_1992,Jacobs_2006,cao_entanglement_2019}\begin{equation}\label{eq:monit}
    \mathcal{D}[\rho]= \g \left( 2\mathcal O \rho \mathcal O - \left\{ \mathcal  O^2 ,\rho \right\} \right) \,.
\end{equation}
We are interested in the particle ($\zeta=0$) and heat ($\zeta=1$) currents flowing into a reservoir $r$, which read
\begin{equation}\label{eq:current}
J^{\zeta}_r=i\sum_{k,i} (\varepsilon_{r,k}-\mu_r)^{\zeta} \left[t_{r,ki}^{*} \langle d_i^{\dag} c^{\phantom \dagger}_{r,k} \rangle-t_{r,ki} \langle c^{\dagger}_{r,k} d^{\phantom \dagger}_i\rangle \right].
\end{equation}
When single-particle observables $\mathcal O=\sum_{ij}d_i^{\dag}O^{\phantom \dag}_{ij}d^{\phantom \dag}_j$ are monitored, calculating Eq.~\eqref{eq:current} becomes a difficult task, since Eq.~\eqref{eq:monit} is non-quadratic.~Even though, for quadratic Hamiltonians, correlation functions obey closed systems of equations~\cite{haken1973exactly,Bernard_2018,BernardJinQSSEP_2019}, efficient numerical calculations can be performed only for finite systems~\cite{turkeshi_2021,Jin_2022,turkeshi2023density}.
We show now that analytical solutions can be obtained with infinite reservoirs thanks to the validity of the self-consistent Born scheme for 2-point correlation functions, extensively discussed in Refs.~\cite{Dolgirev_2020,Jin_2022} and in the Supplemental Material (SM)~\footnote{See the Supplemental Material for technical details and additional information.}.

We consider the retarded, advanced, and Keldysh Green's
functions: $\mathcal{G}^{R}_{ij}(t,t')=-i\theta(t-t')\langle\{d^{\phantom \dagger}_i(t),d_j^{\dagger}(t')\}\rangle$,
$\mathcal{G}_{ij}^{A}(t,t')=[\mathcal{G}_{ji}^{R}(t',t)]^{*}$
and $\mathcal{G}_{ij}^{K}(t,t')=-i\langle[d^{\phantom \dagger}_i(t),d_j^{\dagger}(t')]\rangle$, that we  collect in the matrix $\boldsymbol{    \mathcal{G}}=\bigl( \begin{smallmatrix}\mathcal{G}^R & \mathcal{G}^K\\ 0 & \mathcal{G}^A\end{smallmatrix}\bigr)$~\cite{Kamenev_2011}. 
The matrix $\boldsymbol{\mathcal G}$ obeys Dyson's equation $\boldsymbol{{\cal G}}^{-1}= \boldsymbol{\mathcal G}_0^{-1}-\boldsymbol{\Sigma}$, where $\boldsymbol{\mathcal G}_{0}$ is the Green's function of the isolated system ($t_{r,ki}=\gamma=0$) and $\boldsymbol{\Sigma}$ is the self-energy, encoding the effects of reservoirs and monitoring. 
The contribution of the reservoir $r$ to $\boldsymbol{\Sigma}$ is obtained by integration of the modes $c_{r,k}$. In frequency space, $\boldsymbol{\Sigma}_{r,ij}(\omega)=\sum_{k}t^*_{r,ki}t^{\phantom *}_{r,kj}\boldsymbol{\mathcal C}_{r,k}(\omega)$, where $\boldsymbol{\mathcal C}_{r,k}$ is the Green's functions of the reservoir. Particle exchange with the system is described by the hybridization matrix 
$\Gamma_{r}(\w)=[\S^A_r(\w )-\S^R_r(\w )]/2$~\cite{Datta_1995}.
The Keldysh component $\Sigma^K_r(\omega)=-2i\Gamma_r(\omega)\tanh[(\omega-\mu_r)/2T_r]$  carries information about the equilibrium  state of the reservoirs.  

Monitoring contributes to the self-energy following the self-consistent Born scheme~\cite{Dolgirev_2020,Jin_2022}, which involves the full Green's matrix $\boldsymbol{{\cal G}}$, including baths and monitoring:
\begin{equation}\label{eq:self}
\boldsymbol{\Sigma}^\gamma_{ij}(\w)=2\g\sum_{pq} O_{ip}\boldsymbol{{\cal G}}_{pq}(t,t) O_{qj}\,.
\end{equation}
To derive the retarded and advanced components of $\boldsymbol\Sigma^\gamma$, we exploit the prescription $\mathcal G^{R/A}_{ij}(t,t)=\mp i\delta_{ij}/2$~\cite{Kamenev_2011}, and obtain
$[\mathcal G^{R/A}]^{-1}_{ij}(\omega)=\omega-h_{ij}-\sum_r\Sigma_{r,ij}^{R/A}(\omega)\pm i\gamma \sum_pO_{ip}O_{pj}$. In this expression, monitoring appears as a frequency-independent  life-time  $\gamma \sum_pO_{ip}O_{pj}$, in analogy with  single-particle gains or losses~\cite{corman_dissipativeQPC_2019,Visuri_2022_sym,Visuri_2022,huang2023superfluid,visuri2023dc,jin_generic_2020,PhysRevLett.122.040402,PhysRevB.104.155431}.

The difference between monitoring and losses appears in the Keldysh component of Eq.~\eqref{eq:self}. Inserting $\mathcal G_{ij}^K(t,t)=2i\langle d^\dagger_jd^{\phantom \dagger}_i\rangle-i\delta_{ij}$~\footnote{As we consider the stationary regime,  $\langle d^\dagger_j(t)d^{\phantom \dagger}_i(t)\rangle$ does not depend on time.} and inverting the Dyson equation, one finds a self-consistent equation
for the correlation matrix $\mathcal D_{ij}=\langle d^\dagger_jd^{\phantom \dagger}_i\rangle$ 
\begin{equation}\label{eq:Ds}
    \mathcal  D=\int\frac{d\w}{\pi} \mathcal{G}^R(\omega) \left[\sum_{r}f_r(\omega)\G_r(\omega)+\gamma O\mathcal DO\right]\mathcal{G}^A(\omega) \,.
\end{equation}

The solution of Eq.~\eqref{eq:Ds} completes the full derivation of the Green's function  $\boldsymbol{\mathcal G}$. 
The knowledge of $\boldsymbol{\mathcal G}$ is sufficient to derive currents
~\cite{Meir_1992,jin_generic_2020}. 
After straightforward algebra, detailed in the SM~\footnotemark[2], we 
find closed, exact and non-perturbative expressions for the particle and heat currents:
\begin{align}\label{main}
&J^{\zeta}_r=\frac{2}{\pi} \int d\w\,\ubcolor{ \left( \w -\mu_{r} \right)^{\zeta} \left( f_{\bar{r}} - f_{r} \right) \tr \left[ \G_r {\cal G}^R \G_{\bar{r}} {\cal G}^A \right] }{\text{elastic}} \nonumber \\
&+\gamma\frac2\pi \int d\omega\,\ubcolor{\left( \w -\mu_{r} \right)^{\zeta} \tr \left[ \G_r {\cal G}^R O\left(\mathcal D-f_r \mathbb 1\right)O {\cal G}^A \right] }{\text{inelastic}}  \; .
\end{align}
with $\bar r=R$ if $r=L$ and viceversa. This expression is the main result of our work. It allows us to draw general conclusions on monitor-assisted transport and, combined with Eq.~\eqref{eq:Ds}, can be directly applied to all settings described by Lindbladians of the form~\eqref{eq:monit}. 

Equation~\eqref{main} appears as a sum of two distinct  terms. The first term reproduces the Landauer-B\"uttiker formula for currents in non-interacting systems~\cite{landauer_1957,Datta_1995}. It describes the energy-preserving transfer of particles between reservoirs at energy $\w$  with transmission probability $\mathcal{T}(\w)=4
\tr \left[ \G_r {\cal G}^R \G_{\bar{r}} {\cal G}^A \right]$. 
As $\mathcal T(\omega)$ depends  on  $\mathcal G^{R/A}$, where measurements only contribute to reducing life-times,  monitoring affects elastic transport exactly as single-particle gains or losses~\cite{corman_dissipativeQPC_2019,Visuri_2022_sym,Visuri_2022,huang2023superfluid,visuri2023dc,jin_generic_2020,PhysRevLett.122.040402,PhysRevB.104.155431}.

The second term in Eq.~\eqref{main} is controlled by monitoring. The implicit dependence of the correlation matrix $\mathcal D$  on additional energy integrals, see Eq.~\eqref{eq:Ds}, indicates that measurements inelastically  add or subtract energy to particles in the system. A rough inspection of Eq.~\eqref{main} shows that the inelastic contribution has a peak as function of the observation rate $\g$, interpolating between a linear growth for small $\g$ and a $\g^{-1}$ decay in the strong measurement limit, as $\mathcal G^{R/A}\propto \gamma^{-1}$ for $\gamma\rightarrow\infty$, see Figs.~\ref{fig:setup}b-\ref{fig:model_qssep}c. Position and strength of this maximum depend on the details of the problem,  but it is generally expected for values of $\gamma$ comparable to the spectral width of the system and its coupling strength to the baths. These maxima are out of reach in perturbative approaches. 
Importantly, the inelastic current is not directly proportional to $f_L-f_R$, and can thus be finite even without a bias. This  mechanism describes the generation of non-reciprocal currents from measurement and can be exploited for work generation. 

We provide below explicit illustrations of these considerations on two different monitor-assisted devices.

{\it Monitored density engine -- \label{sec:Unbiased}} We first consider a monitored setting, sketched in Fig.~\ref{fig:setup}, where a single level of energy $\varepsilon_d$, described by the Hamiltonian $\mathcal H_{\rm sys}=\varepsilon_d\, d^{\dag}d$,  evolves under the continuous measurement of its occupation, associated with the operator $\mathcal O=n=d^\dagger d$. 
Solving Eq.~\eqref{eq:Ds} gives  the occupation of the level
\begin{equation}\label{eq:n}
\av n =\frac{\int d\w\, \mathcal A(\omega) [f_{L}(\omega) P_{L}(\omega)+f_{R}(\omega)P_{R}(\omega)]}{\int d\w\, \mathcal A(\omega) [P_{L}(\omega)+P_{R}(\omega)]}\;,
\end{equation}
where $\mathcal A(\w)=-\im [\mathcal G_{dd}^R(\omega)]/\pi=\frac{1}{\pi}\frac{\G_{L}+\G_{R}+\g}{|\w-\varepsilon_d-
\S_{L}-\S_{R}+i\g|^{2}}$  is the spectral function of the level. We have introduced the quantity $P_{r}(\omega)=\G_r/[\G_{\rm L}+\G_{\rm R}+\g]$, 
which highlights the non-equilibrium effects of monitoring. For instance, in the unbiased case ($f_{L,R}=f)$, the absence of dephasing ($\gamma=0$) is needed to recover $P_R+P_L=1$ and the standard equilibrium expression $\av n_{\rm eq}=\int d\omega\,\mathcal A(\omega)f(\omega)$~\footnote{Remarkably, it is also possible to have $\av n=\av n_{\rm eq}$ with  $\gamma\neq0$, provided that both the hybridization functions $\Gamma_r$ do not depend on frequency, which is also a condition to suppress the non-reciprocal current.}. Injecting Eq.~\eqref{eq:n} in Eq.~\eqref{main}, we obtain the particle current $J^0=J^0_R=-J^0_L$ flowing through the system
\begin{align} \label{eq:density} 
&J^{0}=2\int d\w \mathcal A \frac{\G_{L}\G_{R}}{\G_L +\G_{R} +\g}\left(f_L-f_R\right)\\
&+\frac{2\g}{\int d\w \mathcal A\left(P_L+P_{R}\right)}\int d\w d\w'\mathcal A^{\phantom '}\mathcal A' P^{\phantom '}_{L}P'_R\left(f^{\phantom '}_{L}-f'_R\right)\,,\nonumber
\end{align}
where we omit all frequency dependency
for compactness and  use the shorthand notation $f'=f(\omega')$. 

The first term reproduces the well-known expression of the current flowing through a Breit-Wigner resonance~\cite{breitwigner,buttiker_1988_BW}, with an additional suppression controlled by the monitoring rate $\gamma$. 

The inspection of the inelastic term in Eq.~\eqref{eq:density} directly shows that even without bias ($\mu_{L,R}=\mu,\, T_{R,L}=T$), monitoring can trigger the flow of a finite, non-reciprocal current through the system. This non-reciprocal current is finite provided that at least one of the hybridization functions $\Gamma_{L/R}$ depends on energy, and that mirror and particle-hole symmetry are  simultaneously broken~\cite{PhysRevB.83.085428,Sanchez_2017,PhysRevB.95.235404}. Such conditions are satisfied when $\Gamma_{L}\neq\Gamma_R$ and at least one function among $\mathcal A$ or $\Gamma_{L/R}$ is not symmetric around the  chemical potential $\mu$.  The mechanism generating this current is sketched in Fig.~\ref{fig:setup}a: electrons at energy $\w_1$ are emitted from one reservoir onto the level and the measurement provides the energy for the electron to exit into an empty state of the other reservoir at energy $\omega_2$. The fact that the injection and emission rates depend asymmetrically on energy allows the generation of the current. 
The emergence of a non-reciprocal current can be also understood based on  the fact that averaging over the measurement outcomes is equivalent, in this specific case, to coupling the system to an infinite-temperature bosonic bath (see SM~\footnotemark[2]), which induces a thermoelectric  flow in the system if mirror and particle-hole symmetry are broken~\cite{Mazza_2014,Sothmann_2014,Fleury_2021}. 

Figure~\ref{fig:setup}b shows that the inelastic current displays the aforementioned peak as a function of the measurement strength $\gamma$ at zero bias $\delta\mu=\mu_L-\mu_R=0$. For all numerical applications, we consider a minimal model  where the level is coupled to two metallic reservoirs via two energy filters of energy $\varepsilon_{L/R}$. In this case, $\S^R_{r}(\omega)=t^2/(\w-\varepsilon_{r}+i \Delta)$, where $t$ is the level-filter tunnel coupling and $\Delta$ the hybridization constant of the filter with  the reservoirs, see SM~\footnotemark[2]. The resulting hybridization function $\Gamma_r(\omega)=-\mbox {Im}\Sigma^R_r(\omega)$ is peaked around $\varepsilon_r$, as sketched in Fig.~\ref{fig:setup}a. We have found the maximum non-reciprocal current  for $\gamma\simeq t$ -- that is out of weak coupling ($\gamma\gg t$) -- when  $\varepsilon_d=0$ and  when mirror and particle-hole symmetry are broken by  antisymmetric reservoirs with $\varepsilon_L=-\varepsilon_R$.  The peak roughly follows $\varepsilon_d$ and is suppressed by finite  temperatures, see inset of Fig.~\ref{fig:anomalous}b. Similar non-reciprocal effects and peaks were also discussed, from a real-time perspective, in Refs.~\cite{Bredol_2021_alone,Bredol_2021}.

Figure~\ref{fig:setup}c  shows the differential  conductance  $G=\partial J^0/\partial\delta\mu|_{\delta\mu=0}$ for the same system. $G$ also features elastic and inelastic contributions~\cite{pastawski1991,pastawski1992}, scaling differently with $\g$. For small rates, the elastic term dominates, 
showing as many peaks as resonances in the system -- three in the application of Fig.~\ref{fig:setup}. Because of the fact that  only the central level is monitored, increasing $\gamma$ strongly suppresses its associated resonance, while spectral weight is transferred to the filters (arrows in Fig.~\ref{fig:setup}c). Consequently, for intermediate  monitoring strengths $\gamma\simeq t$,  the conductance actually increases out of resonance ($\mu\neq0$), before being also suppressed in the $\gamma\gg t$ limit.

The fact that monitoring generates currents at zero bias, implies that they can flow against externally imposed biases to generate work. We consider here the generated power $\mathcal P=\delta\mu\cdot J^0$ and show the importance of non-perturbative and out-of-equilibrium effects on this quantity. 
In linear response, $J^0\simeq J^0|_{\delta\mu=0}-\delta\mu \,G$, and the power has a parabolic dependence on $\delta\mu$,   with a maximum  $\mathcal P_{\rm max}=J^{0}|_{\delta\mu=0}^2/2G$ and a change of sign at the stopping voltage $\delta\mu_{\rm stop}=J^0|_{\delta\mu=0}/G$.
Figure~\ref{fig:setup}d  shows that the maximum power generation is found for monitoring of strength $\gamma>t$, that is out of the weak coupling regime.  Moreover, we find that non-equilibrium effects associated to strongly biased reservoirs cannot be neglected. They can be exactly derived via Eq.~\eqref{eq:density}, and the dashed lines in Fig.~\ref{fig:setup}d clearly show that  linear-response greatly overestimates $\mathcal P_{\rm max}$ and  $\delta\mu_{\rm stop}$  when $\gamma\simeq t$.

{\it Quantum measurement cooling -- }We consider  two independent sites $H_{\text{sys}}=\varepsilon^{\phantom \dagger}_L d^{\dag}_L d^{\phantom\dag}_L + \varepsilon^{\phantom \dagger}_R d^{\dag}_R d^{\phantom\dag}_R$  that are coupled via the monitoring process, $O_{ij}=\d_{iL}\d_{jR}+\d_{iR}\d_{jL}$, see Fig.~\ref{fig:model_qssep}a. This process can be in principle realized by adding an interferometer measuring cross-correlations between the two sites~\cite{Gullans_2019,hruza_2023}.  
Also in this case, we rely on Eq.~\eqref{eq:Ds} to find the occupation of the levels $\av {n_r}=\av{d^\dagger_r d^{\phantom\dagger}_r}$ \begin{equation}\label{eq:nr}
\av{n_r}=\frac{\int d\omega \big[f_{\bar r}P_{\bar r}\mathcal A_{\bar r}+\big(1-\int d\omega' P'_{\bar r}\mathcal A'_{\bar r}\big)f_{r}P_{r}\mathcal A_r\big]}{\sum_{r'}\int d\omega P_{r'}\mathcal A_{r'}-\prod_{r'}\int d\omega P_{r'}\mathcal A_{r'}}\,,
\end{equation}
with modified notation $P_r=\G_r/(\G_r+\g)$ and spectral functions $\mathcal A_r(\w)=-
\mbox{Im}\mathcal G^R_{rr}/\pi=\frac{1}{\pi}\frac{\G_{r}+\g}{|\w-\varepsilon_r-
\S_{r}+i\g|^{2}}$. 

Because of the absence of coherent hopping between sites, $\mathcal G^{R/A}_{LR}=0$ and only the inelastic component of the currents in Eq.~\eqref{main} is finite, for which the knowledge of Eq.~\eqref{eq:nr} is needed. We are interested in exact expressions for quantum measurement cooling (QmC)~\cite{buffoni_QMC_2019},  
we thus consider the heat current flowing in the right reservoir 

\begin{multline}\label{double}
J^{1}_{R}=\frac{2\g}{\mathcal N} \int d\w d\w' (\omega-\mu_R)\mathcal A_{R}P_{R} \Big[\mathcal A'_{L}P'_{L}\left(f'_{L}-f_R\right)~~~\\
~~~~~~~~~~~+ \Big(1- \int d\w'' \mathcal A''_{L} P''_{L} \Big) \mathcal A'_{R}P'_{R}\left(f'_{ R}-f_{R}\right)\Big] \; ,
\end{multline}
where $\mathcal N$ is the denominator appearing in Eq.~\eqref{eq:nr}. To get physical insight on the physical requirements for QmC and the multiple processes described by Eq.~\eqref{double}, we first inspect  the $\gamma\rightarrow 0$ limit.  To leading order in $\gamma$, one can approximate $P_r=1$ and only the first term in Eq.~\eqref{double} remains. It can be cast in the compact form 
\begin{equation}
\begin{split}
J^1_R=&2\gamma\int d\omega\, (\omega-\mu_R)\mathcal{A}_R(\omega)\big[\av{n_{L}}-f_R(\omega)\big]\,.
\end{split}
\end{equation}
If we further approximate the spectral function by $\mathcal A_R(\omega)=\delta(\omega-\varepsilon_R)$, we  get $J^1_R=2\gamma (\varepsilon_R-\mu_R)(\av{n_{L}}-\av{n_R})$. This expression makes explicit that the heat flow in the right reservoir is controlled by the position of the right level with respect to the chemical potential and the difference of occupation with respect to the left level.  The condition for cooling the right reservoir is $J^1_R<0$. In the absence of bias, such condition requires $\mu_R\lessgtr\varepsilon_R$ and $\varepsilon_{L}\lessgtr\varepsilon_R$, as sketched in Fig.~\ref{fig:model_qssep}a.  Analogous conditions were found to achieve cooling by heating~\cite{cleuren_CBH_2012,mari_CBH_2012}, where the role of measurement is played by a third hot reservoir. 
The second term in Eq.~\eqref{double} acts at order $\gamma^2$ and describes the reinjection of heat in the right reservoir by particles hopping back and forth to the left level via the monitoring process.

In Figure~\ref{fig:model_qssep}, we explore QmC and its performances also for strong temperature biases and large values of $\gamma$. For numerical applications, we consider $\mu_{L/R}=0$ and take the same  hybridization functions $\Gamma_r(\omega)$ than in the previous section, with peaks aligned with $\varepsilon_r$.
Figure~\ref{fig:model_qssep}b shows the regions where QmC occurs, in the absence of bias and for increasing monitoring strength $\gamma$. QmC indeed occurs when $\varepsilon_L\lessgtr\varepsilon_R \lessgtr0$. Nonetheless, the parameter region for QmC shrinks the larger the monitoring strength $\gamma$, reflecting the fact that more heat is injected in both reservoirs the stronger the measurement process is. 
Fig.~\ref{fig:model_qssep}c shows the behavior of $J^1_R$ for increasing temperature biases as function of $\gamma$. Exactly as the non-reciprocal current discussed in the previous section (Fig.~\ref{fig:setup}b), the heat current shows a peak for $\gamma\approx t$. However, increasing the temperature bias leads to a change of sign of the heat current, signaling that the left reservoir is hot enough to heat the right one. 

We conclude this study by discussing the efficiency of this process, which is characterized by the coefficient of performance, $COP=|J^1_R/(J^1_R+J^1_L)|$, which measures how much heat can be extracted from monitoring~\cite{callen1998thermodynamics}. We depict the COP in Fig.~\ref{fig:cop}d as a parametric plot on the rate $\g$. For fixed temperatures in the reservoirs, the maximum COP is found near the critical measurement strength $\gamma$ at which the heat flow changes sign in Fig.~\ref{fig:cbh_region}c.  This monitoring  strength is also of order $t$ and is not encompassed by the weak coupling limit.

{\it Conclusions -- }We have derived exact and analytic expressions for the particle and heat currents flowing in a large class of monitored systems. These formulas were applied to investigate power harvesting and cooling assisted by measurements. In particular, we have found current peaks as a function of the measurement strength $\gamma$  out of the weak-coupling limit (Figs. \ref{fig:setup}b-\ref{fig:cbh_region}c).  These peaks are clear features that could favor their observation in experiments. Our results can be readily generalized to different monitored setups and pave the way to the investigation of unexplored regimes which are not captured by standard, perturbative  approaches. We have shown that these regimes are important, as they manifest the best performances in terms of power generation and quantum measurement cooling. 

On a more fundamental level, we have provided exact expressions for quantum transport in the presence of non-elastic effects caused by monitoring. It would be of great interest to establish in the future whether formulas like Eq.~\eqref{main} also apply for interacting quantum impurity models driven out of equilibrium, and/or  for systems coupled to bosonic baths at finite or even zero temperature~\cite{Glazman_1988,Wingreen_1988,PhysRevB.50.5528,PhysRevB.70.075301,braak2020fermi}.

\begin{acknowledgments}
{\it Acknowledgments  -- } We are grateful to Christophe Berthod, Daniel Braak, G\'eraldine Haack, Manuel Houzet, Rafael S\'anchez, Kyrylo Snizhko, Clemens Winkelmann and Robert Whitney for helpful comments and discussions. This work has been supported by the Swiss National Science Foundation under Division II under grant 200020-188687. J.S.F. and M.F. acknowledge support from the FNS/SNF Ambizione Grant No. PZ00P2\_174038. M. F. acknowledges support from  EPiQ ANR-22-PETQ-0007 part of Plan France 2030.
\end{acknowledgments}

\bibliographystyle{apsrev4-1}
\bibliography{biblio}

\newpage

\onecolumngrid

\setcounter{equation}{0}
\setcounter{figure}{0}
\setcounter{table}{0}
\setcounter{page}{1}
\makeatletter
\renewcommand{\theequation}{S\arabic{equation}}
\renewcommand{\thefigure}{S\arabic{figure}}
\renewcommand{\bibnumfmt}[1]{[#1]}
\renewcommand{\citenumfont}[1]{#1}

\pagebreak
\begin{center}
    {\Large \bf Supplemental Material of ``Exact description of transport and non-reciprocity in monitored quantum devices'' \par}
\vspace{0.1cm}
\end{center}

\small

In this Supplemental Material, we recall how averaging over the results of weak measurements of an operator $\mathcal O$ leads to the Lindblad dynamics described by Eq.~(1) in the main text. We then derive explicitly the contributions to the self-energy $\boldsymbol\Sigma$ coming from non-interacting reservoirs with energy filters, used for numerical applications in the main text,  and, in particular, from monitoring -- Eq.~(3) in the main text. We provide details about the derivation of the self-consistent formula for the correlation matrix $\mathcal D$ -- Eq.~(4) in the main text -- and the exact expressions for the particle and heat currents -- Eq.~(5) in the main text. We then establish the equivalence between the Lindblad dynamics induced by averaged measurements of the operator $\mathcal O$ and the coupling of the operator $\mathcal O$ to a bosonic bath at infinite temperature and large chemical potential. 

\normalsize
\vspace{1cm}

\setcounter{equation}{0}
\setcounter{figure}{0}
\setcounter{table}{0}
\setcounter{page}{1}
\makeatletter
\renewcommand{\theequation}{S\arabic{equation}}
\renewcommand{\thefigure}{S\arabic{figure}}
\renewcommand{\bibnumfmt}[1]{[#1]}
\renewcommand{\citenumfont}[1]{#1}


\onecolumngrid

\section{Lindblad dynamics induced in average by continuous monitoring }

\subsection{Continuous weak monitoring}

In this section, we briefly discuss how the Lindbladian in Eq.~(1) of the main text originates from an evolution under continuous  measurement of the operator $\mathcal{O}=\sum_{ij}d^\dag_i O^{\phantom\dag}_{ij} d^{\phantom\dag}_j$ with $O_{ij}^*=O_{ji}$. The associated monitored dynamics of the quantum state is described by a stochastic Schr\"{o}dinger equation. Taking the It\=o prescription, the infinitesimal increment of the wave-function $d|\psi_t\rangle=|\psi_{t+dt}\rangle-|\psi_t\rangle$ obeys~\cite{Jacobs_2006} 
\begin{equation}
    d \ket{\psi_t}=\left[-i\mathcal H  -\g \left(\mathcal{O}-\av{\mathcal{O}}_t \right)^2 \right] \ket{\psi_t} dt + \sqrt{2\g} \left(\mathcal{O}-\av{\mathcal{O}}_t \right)  \ket{\psi_t} dB_t \;,
\end{equation}
where $\mathcal H$ is the Hamiltonian of the system, $\langle\mathcal O\rangle_t=\langle\psi_t|\mathcal O|\psi_t\rangle $  and $\gamma$ is the measurement strength. The quantity $dB_t$ is the increment of a stochastic Wiener process $B_t$, which, according to It\=o rules~\cite{Oeksendal_2003}, behaves in average as $\mathbb{E}[dB_t] = 0$ and $dB_t^2 = dt$. 
The equation ruling the dynamics of the mean density matrix $\rho_t=\mathbb{E}\left[\ket{\psi_t}\bra{\psi_t} \right]$ is then obtained by averaging over all measurement outcomes and over the noise realizations $B_t$. It satisfies the differential equation
\begin{align}
    d\rho_t&=\mathbb{E}\left[(d\ket{\psi_t})\bra{\psi_t}+\ket{\psi_t}(d\bra{\psi_t}) + (d\ket{\psi_t})(d\bra{\psi_t})\right] \\
    &= \mathbb{E}\left[ -i[\mathcal H,\ket{\psi_t}\bra{\psi_t}] dt-\g [\mathcal{O},[\mathcal{O},\ket{\psi_t}\bra{\psi_t}]] dt + \sqrt{2\g} \{\mathcal{O}-\langle\mathcal O\rangle_t,\ket{\psi_t}\bra{\psi_t}\} dB_t \right]  \\
    &= -i[\mathcal H,\rho_t] dt -\g [\mathcal{O},[\mathcal{O},\rho_t]] dt  \;,\label{eq:meas}
\end{align}
where we have neglected all terms of order $dB_tdt$ and $dt^2$ and used the  It\={o} rules  $dB_t^2 = dt$ and $\mathbb{E}[dB_t] = 0$. 

Importantly, we see that the second term in Equation~\ref{eq:meas} involves a double commutator, which reproduces the Lindblad form $\mathcal D[\rho]=\gamma(2\mathcal O\rho\mathcal O-\{\mathcal O^2,\rho\})$, see also Eq.~(1) in the main text. Since we consider observables $\cal{O}$ and Hamiltonians $\mathcal H$ that are quadratic in the creation and annihilation operators,  on average, the contribution of the measurements to the evolution equation of $n$-point correlators close on themselves. 

We illustrate this last point on the two-point function $G_{k,l}(t)=\langle d^{\phantom\dagger}_kd^\dagger_l\rangle^{\phantom\dagger}_t={\rm tr}[\rho^{\phantom\dagger}_{t}d^{\phantom\dagger}_{k}d_{l}^{\dagger}]$. If we suppose a quadratic Hamiltonian  of
the form $\mathcal H=\sum_{i,j}d_{i}^{\dagger}h^{\phantom \dagger}_{ij}d^{\phantom \dagger}_{j}$, thanks to the cyclic properties of the trace, one finds that the time evolution of the correlation function $G$ is ruled by the differential equation
\begin{equation}
\frac{d}{dt}G_{k,l}(t)=\frac{d}{dt}\langle d^{\phantom \dagger}_kd^\dagger_l\rangle^{\phantom\dagger}_t=i\langle[\mathcal H,d^{\phantom \dagger}_kd^\dagger_l]\rangle^{\phantom\dagger}_t-\gamma\langle[\mathcal O,[\mathcal O,d^{\phantom \dagger}_kd^\dagger_l]]\rangle^{\phantom\dagger}_t\,.
\end{equation} 
Since both the operators $\mathcal H$ and $\mathcal O$ are quadratic in the creation and annihilation operators $d$ and $d^\dagger$, the commutators also generate quadratic operators. As a consequence, one can write down a closed system of equations for the elements of the correlation matrix $G$:
\begin{equation}
\frac{d}{dt}G(t)=-i [h,G(t)]-\gamma[O,[O,G(t)]].
\end{equation} 
Such considerations extend to any $n$-point correlation function.


\subsection{Equivalence to averaged evolution of a quantum stochastic Hamiltonian}     
Alternatively, the averaged evolution equation~\eqref{eq:meas} can also be generated
by the quantum stochastic Hamiltonian (QSH): 
\begin{equation}\label{eq:stochHamiltonian}
d\mathcal H_{t}=\sqrt{2\gamma}OdB_{t},
\end{equation}
where $B_{t}$ is the same Wiener process as above. The infinitesimal evolution of the wave-function is unitary and reads  
\begin{align}
\left|\psi_{t+dt}\right\rangle& =e^{-i\mathcal Hdt-id\mathcal H_{t}}\left|\psi_{t}\right\rangle &\Longrightarrow&& d|\psi_t\rangle&=-i(\mathcal H+d\mathcal H_t)dt|\psi_t\rangle -\gamma\mathcal O^2 dt|\psi_t\rangle \,,
\end{align}
where we have again discarded terms of order $dB_t dt$ and $dt^2$ and applied $dB_t^2=dt$.  The infinitesimal
evolution generated by this term on the density matrix $\rho_{t}=\mathbb{E}\left[\left|\psi_{t}\right\rangle \left\langle \psi_{t}\right|\right]$
reads
\begin{align}
d\rho_{t} & =-i[\mathcal H,\rho_{t}]dt-\frac{1}{2}[d\mathcal H_{t},[d\mathcal H_{t},\rho_{t}]] =-i[\mathcal H,\rho_{t}]dt-\gamma[\mathcal O,[\mathcal O,\rho_{t}]]dt,
\end{align}
where we have again made use of the fact that $\mathbb{E}[dB_{t}]=0$ and which coincides with Eq.~\eqref{eq:meas}. 

If we are only interested in the average evolution of the density
matrix, it is advantageous to use the unraveling in terms of QSHs. Indeed,
for this procedure, the evolution at the stochastic level is both
unitary and quadratic in fermionic creation and annihilation operators,
meaning that Gaussian states are preserved under evolution. Additionally,  the equations of motion remain linear at the stochastic level - differently from the monitored case. 

Note however that, despite these advantages, the inclusion of thermal baths with finite memory time is still challenging, as it will render the equations of motion non-local in time. 
An additional powerful property of QSHs for that matter is that, at the field theory level, the self-consistent Born approximate (SCBA) is \emph{exact} for QSHs~\cite{Dolgirev_2020,jin_generic_2020}. This allows to express all observables of interest in a closed-form even in the presence of thermal baths as we will discuss in Section~\ref{sec:self_monitoring}.


\section{Self-energy of the system: reservoirs and monitoring contributions}
We recall that we consider  density matrices $\rho$ evolving according to the Lindblad equation 
\begin{align}\label{eq:rhot}
    \partial_t \rho&=-i\left[\mathcal H,\;\rho \right]
    +\g \left( 2\mathcal{O} \rho \mathcal{O} - \{\mathcal{O}^2,\rho\}\right)
    \,, 
\end{align}
where $\mathcal O=\sum_{ij}d_i^{\dag}O^{\phantom \dag}_{ij}d^{\phantom \dag}_j$ is the monitored single particle operator. The Hamiltonian describes a system tunnel coupled to reservoirs in the generic form  $\mathcal H=\mathcal  H_{\rm res}+\mathcal H_{\rm T}+\mathcal H_{\rm sys}$, with 
\begin{equation}\label{eq:ham}
   \mathcal H=\ubcolor{\sum_{k,r=L,R} \varepsilon^{\phantom \dagger}_{r,k}c^{\dag}_{r,k}c^{\phantom \dagger}_{r,k}}{\text{Reservoir}}
    +\ubcolor{\sum_{k,i,r=L/R} \Big[t^{\phantom \dagger}_{r,ki} c^{\dag}_{r,k}d^{\phantom \dagger}_i + t^{*\phantom \dagger}_{r,ki} d_i^{\dag} c^{\phantom \dagger}_{r,k}\Big] }{\text{Tunnel Coupling}} +\ubcolor{\sum_{i,j}  d^\dagger_i h^{\phantom\dagger}_{ij}d^{\phantom\dagger}_j }{\text{System}}\;,
\end{equation}
where $c_{r,k}$ destroys electrons in the mode $k$ of the left ($r=L$) or right ($r=R$) reservoir and  $\{d^{\dag}_i,d^{\phantom \dagger}_i\}$ forms a complete, ortho-normal set of single-electron creation and annihilation operators acting on the system.

To derive the self-energies associated to reservoirs and monitoring, we  rely on the Keldysh path-integral formalism~\cite{Kamenev_2011}, 
which can be extended to dissipative systems~\citep{Sieberer_2016}.  In this formalism, the partition function of the system ${\cal Z}={\rm tr}[\rho(t)]$ is expressed in the form 
\begin{equation}
\mathcal{Z}=\int{\mathcal D}\left[\bar c, c,\bar d, d\right]e^{i\mathcal S\left[\bar c, c,\bar d, d\right]}\;,
\end{equation} 
where $\left[\bar c, c,\bar d, d\right]$ is a set of  Grassmann variables on the Keldysh contour corresponding to the creation  and annihilation operators in Eq.~\eqref{eq:ham}. The action $\mathcal S$ corresponding to Eq.~\eqref{eq:rhot} has a unitary and a dissipative component $\mathcal S=\mathcal S_U+\mathcal S_D$. We adopt Larkin-Ovchinnikov's convention~\citep{Larkin_1977} for generic Grassman variables $\psi$
\begin{align}\label{eq:grassman}
\psi^{1}&=\frac{\psi^{+}+\psi^{-}}{\sqrt2}\,, & \psi^{2}&=\frac{\psi^{+}-\psi^{-}}{\sqrt2}\,,&\bar{\psi}^{1}&=\frac{\bar{\psi}^{+}-\bar{\psi}^{-}}{\sqrt2}\,, &\bar{\psi}^{2}&=\frac{\bar{\psi}^{+}+\bar{\psi}^{-}}{\sqrt2}\,.
\end{align}
In this convention, the unitary action corresponding to the Hamiltonian~\eqref{eq:ham} reads, in frequency space,  
\begin{align}\label{eq:actionU}
\mathcal S_U&=\int \frac{d\omega}{2\pi}\left[\sum_{k,r} \bar c_{r,k}(\omega)\boldsymbol {\mathcal C}^{-1}_{r,k}(\omega)
c_{r,k}(\omega)-\sum_{r,ki}\Big[t^{\phantom *}_{r,ki}\,\bar c^{\phantom *}_{r,k}(\omega)\cdot d^{\phantom *}_i(\omega)+t^*_{r,ki}\,\bar d^{\phantom *}_i(\omega)\cdot c^{\phantom *}_{r,k}(\omega)\Big]
+\sum_{ij}\bar d_i(\omega) \boldsymbol {\mathcal G}^{-1}_{0,ij}(\omega)d_j(\omega)
\right]\,,
\end{align}
where we used the shorthand vector notation $\bar{c}=(\bar c^1\quad\bar c^2)$, $c=(c^1\quad c^2)^T$ and analogous for $d$ and $\bar d$. The Green's functions have a retarded, advanced and Keldysh component
\begin{equation}
\boldsymbol{\mathcal{G}}=\left(\begin{array}{cc}
\mathcal{G}^{ R} & \mathcal{G}^{ K}\\
0 & \mathcal{G}^{ A}
\end{array}\right)\,.
\end{equation}
In the case of the action~\eqref{eq:actionU}, the inverse of the Green's
function of the reservoirs and system read, respectively,
\begin{align}\label{eq:CG0}
\boldsymbol {\mathcal C}^{-1}_{r,k}(\omega)&=\begin{bmatrix}
\omega-\varepsilon_{r,k}+i0^+&2i\pi0^+\tanh\left[\frac{\omega-\mu_r}{2T_r}\right]\\
0&\omega-\varepsilon_{r,k}-i0^+
\end{bmatrix}\,,&
\boldsymbol {\mathcal G}^{-1}_{0,ij}(\omega)&=\begin{bmatrix}
\omega-h_{ij}+i0^+&0\\
0&\omega-h_{ij}-i0^+
\end{bmatrix}\,,
\end{align}
where the infinitesimal Keldysh component of $\boldsymbol{\mathcal G}_0$ can be safely ignored, being regularized by the integration of the reservoirs, see Section~\ref{sec:resevoir}. 

The dissipative contribution of the action is instead diagonal in time and reads
\begin{align}\label{eq:stoca}
\mathcal S_{D}&=i\g \int dt \sum_{ijkl}O_{ij}O_{kl}\Big[
\bar d^1_{i}(t)d^1_{j}(t) \bar d^1_{k}(t) d^1_{l} (t)
+2\bar d^1_{i}(t) d^1_{j}(t) \bar d^2_{k}(t) d^2_{l}(t) 
+\bar d^2_{i}(t) d^2_{j}(t) \bar d^2_{k}(t) d^2_{l}(t) 
\Big]\,.
\end{align}
This action can be derived by averaging over the quantum stochastic unitary dynamics, see Section~\ref{sec:self_monitoring}.

Both the presence of reservoirs and monitoring modify the Green's functions  of the system $\boldsymbol{\mathcal G}$ through a self-energy $\boldsymbol{\S}$, according to the Dyson equation
\begin{equation}\label{eq:DysonApp}
\boldsymbol{\mathcal{G}}=\Big[\boldsymbol{\mathcal{G}}_0^{-1}-\boldsymbol{\Sigma}\Big]^{-1}\,.
\end{equation}
We proceed with the calculation of such contributions below.   

\subsection{Self-energy with generic reservoirs and explicit expression with energy filters}\label{sec:resevoir}
The contribution to the self-energy due to the presence of reservoirs, $\boldsymbol{\S}_{L/R}$, can be obtained by Gaussian integration of the reservoirs, namely of the fields $c_{r,k}$ and $\bar c_{r,k}$ in the action~\eqref{eq:actionU}.  Completing the square in Eq.~\eqref{eq:actionU}, one finds the self-energy associated to generic reservoirs reported in the main text 
\begin{equation}\label{eq:selfr}
\boldsymbol{\Sigma}_{r,ij}(\omega)=\sum_k t^*_{r,ki}t^{\phantom *}_{r,kj}\boldsymbol{\mathcal C}_{r,k}(\omega)=\sum_k t^*_{r,ki}t^{\phantom *}_{r,kj}
\begin{bmatrix}
\frac1{\omega-\varepsilon_{r,k}+i0^+}&-2i\pi\delta(\omega-\varepsilon_{r,k})\tanh\left(\frac{\omega-\mu_r}{2T_r}\right)\\
0&\frac1{\omega-\varepsilon_{r,k}-i0^+}\\
\end{bmatrix}\,.
\end{equation}

\begin{figure}
\begin{centering}
\includegraphics[height=3cm]{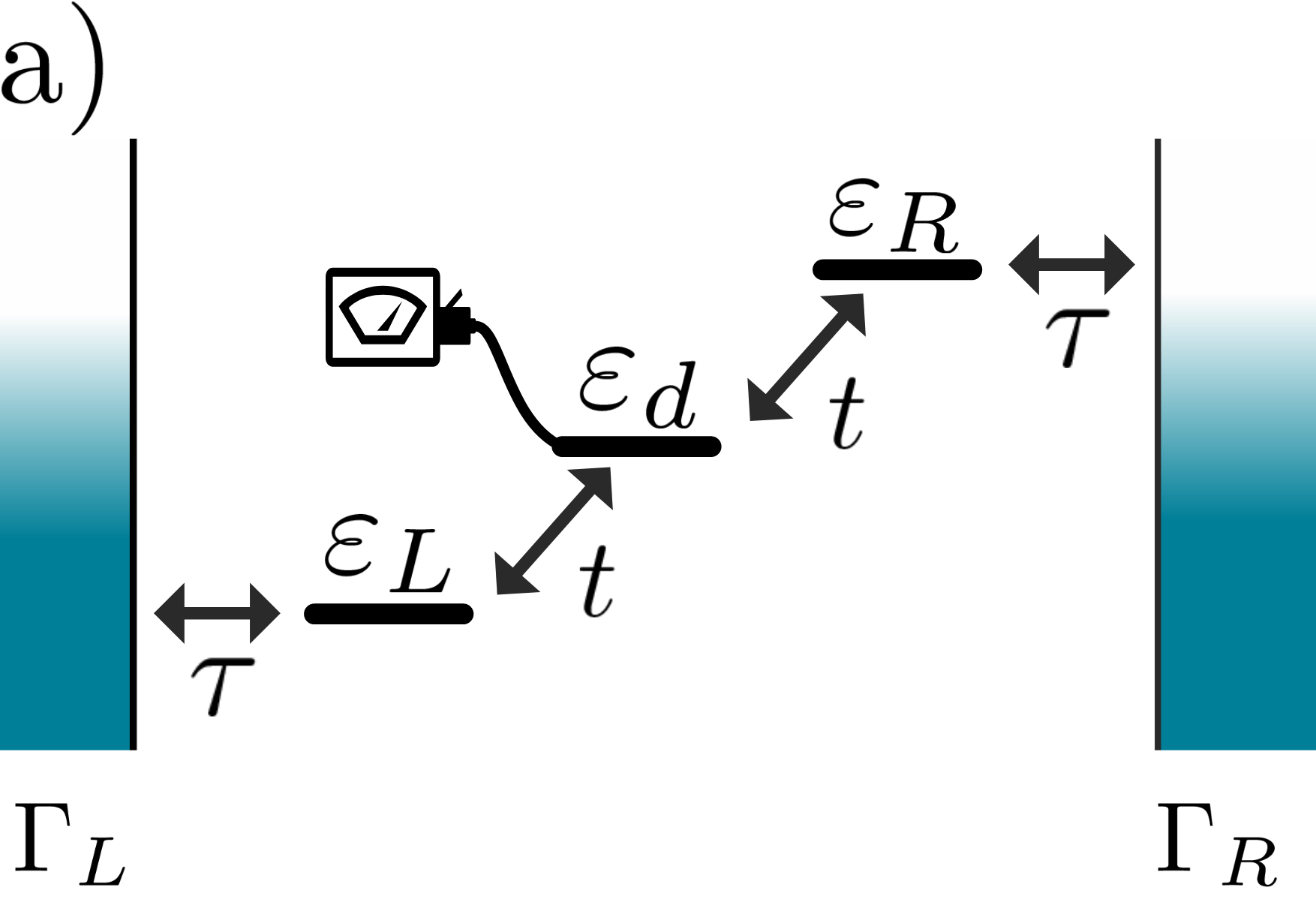}\qquad ~\qquad
\includegraphics[height=3cm]{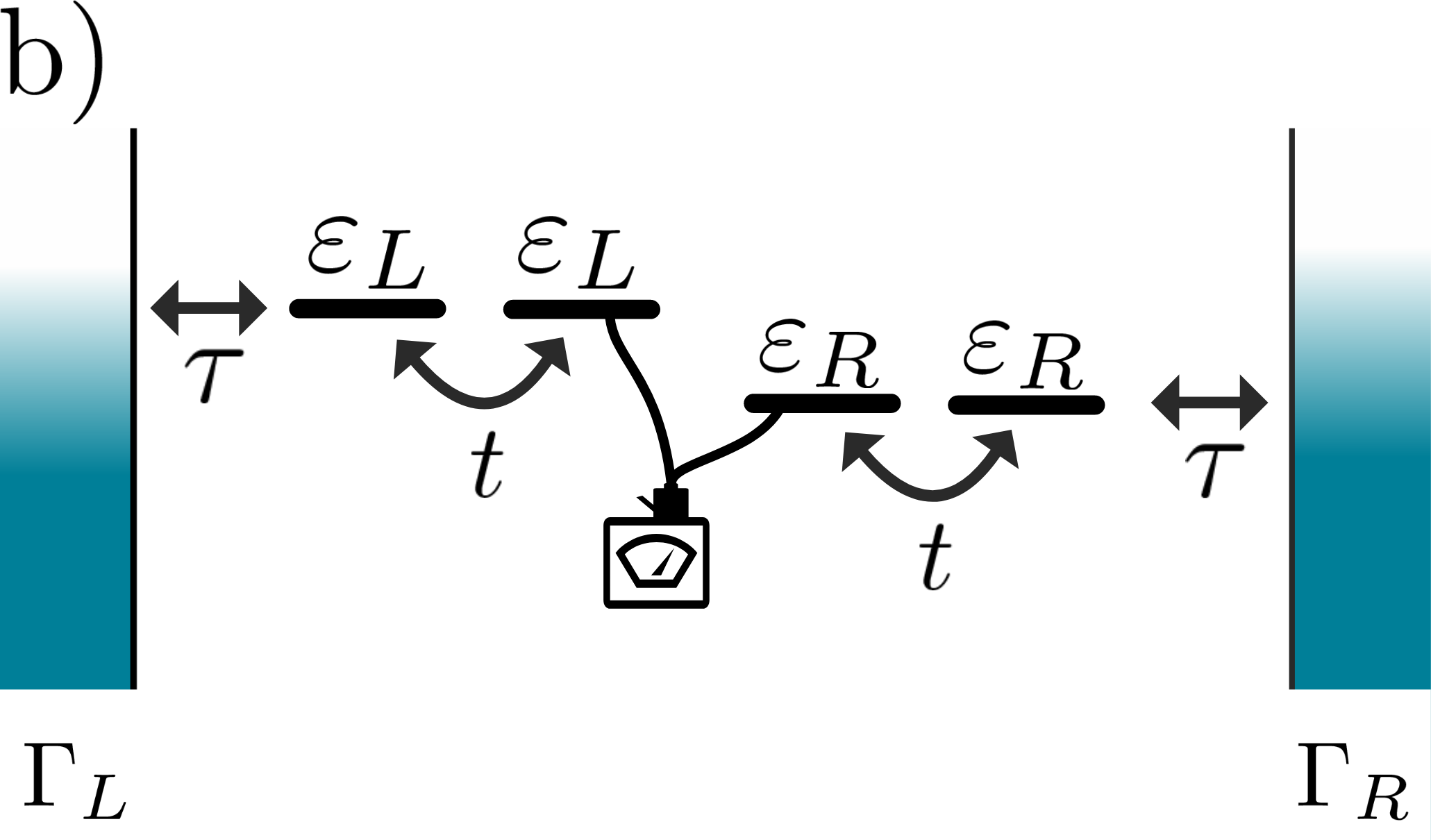}
\par\end{centering}
\caption{Schematic representation of the systems described by the Hamiltonians~\eqref{eq:H1} and~\eqref{eq:H2}, used for numerical applications concerning a) Monitored density engine and b) Quantum measurement cooling in the main text, respectively. Each reservoir is tunnel coupled to a single level through a constant tunnel coupling $\tau$. These levels act as energy filters, which are in turn coupled to the system via a tunneling strength $t$. 
}\label{fig:H-skt}
\end{figure}

We derive an explicit expression of this self-energy for the numerical applications used in the main text. The simplest way to obtain an explicit form of a frequency-dependent hybridization function $\Gamma_r(\omega)$ 
consists in considering  the reservoirs as a lead with a constant density of states $\nu_0$, which is coupled to the system via an energy filter of energy $\varepsilon_r$, see sketches in Fig.~\ref{fig:H-skt}. We also neglect all energy dependence of the tunnel amplitudes. The Hamiltonians describing this situation, and which are used for the two numerical applications in the main text, read  
\begin{align}
\mathcal H_1&=\sum_{r,k}\Big[\varepsilon_{r,k}c^\dagger_{r,k}c^{\phantom\dagger}_{r,k}+\tau \Big(c^\dagger_{r,k}c^{\phantom\dagger}_r+c^\dagger_{r}c^{\phantom\dagger}_{r,k}\Big)\Big]+\sum_{r}\varepsilon_rc^\dagger_rc^{\phantom\dagger}_r+t\sum_r\Big[c^\dagger_r d+d^\dagger c^{\phantom\dagger}_{r}\Big]+\varepsilon_d d^\dagger d\,,\label{eq:H1}\\
\mathcal H_2&=\sum_{r,k}\Big[\varepsilon_{r,k}c^\dagger_{r,k}c^{\phantom\dagger}_{r,k}+\tau \Big(c^\dagger_{r,k}c^{\phantom\dagger}_r+c^\dagger_{r}c^{\phantom\dagger}_{r,k}\Big)\Big]+t\sum_r\Big[c^\dagger_r d^{\phantom \dagger}_r+d^\dagger_r c^{\phantom\dagger}_{r}\Big]+\sum_{r}\varepsilon_r\Big[c^\dagger_rc^{\phantom\dagger}_r+d^\dagger_rd^{\phantom\dagger}_r\Big]\,,\label{eq:H2}
\end{align}
where the operators $c^{\phantom \dagger}_r$ and $c^\dagger_r$ act on the filters and where we have also made explicit, in $\mathcal H_2$, that filters and levels have the same energy $\varepsilon_r$. This is the assumption used in the numerical applications of the main text in the context of quantum measurement cooling, but it is not necessary at all for the following discussion.

Switching back to the field theory language, we proceed now with the integration of all the fields $(c,\,\bar c)$, associated with the reservoirs. We start with those with labels $(r,k)$, corresponding to a metallic lead with a constant density of states $\nu_0$.  Focusing on the case of $\mathcal H_1$ in Eq.~\eqref{eq:H1}, their integration leads to a self-energy contribution of the form~\eqref{eq:selfr} to the energy-filters $(c_r,\bar c_r)$, which leads to a Keldysh action of the form
\begin{multline}\label{eq:SF}
\mathcal S_1=\int \frac{d\omega}{2\pi} \left\{\sum_r \bar c_r(\omega) \begin{bmatrix}\omega-\varepsilon_r+i\Delta&2i\Delta\tanh\left(\frac{\omega-\mu_r}{2T_r}\right)\\
0&\omega-\varepsilon_r-i\Delta\end{bmatrix}c_r(\omega)\right.\\\left.
-t\sum_{r}\Big[\bar c^{\phantom *}_{r}(\omega)\cdot d(\omega)+\bar d(\omega)\cdot c^{\phantom *}_{r}(\omega)\Big]
+\bar d(\omega) \begin{bmatrix}\omega-\varepsilon_d&0\\0&\omega-\varepsilon_d\end{bmatrix}d(\omega)\right\}\,,
\end{multline}
where $\Delta=\pi\nu_0\tau^2$ is the standard expression of the hybridization constant for a single level coupled to a metallic reservoir with a constant density of states in the wide-band limit. The action corresponding to $\mathcal H_2$ in Eq.~\eqref{eq:H2} is analogous. Notice also that the Keldysh component of the inverse Green's function in the first term of Eq.~\eqref{eq:SF} is now finite and we can send the infinitesimal term in the Keldysh component of the last term of Eq.~\eqref{eq:SF} safely to zero. The Gaussian integration of the filters $(c_r,\bar c_r)$ leads to the final form of the self-energy used for the numerical applications in the main text, namely
\begin{equation}\label{eq:selfres}
\Sigma_r(\omega)=\begin{bmatrix}
\frac {t^2}{\omega- \varepsilon_r+i\Delta}&-2i\Gamma_r(\omega)\tanh\left(\frac{\omega-\mu_r}{2T_r}\right)\\
0&\frac {t^2}{\omega- \varepsilon_r-i\Delta}
\end{bmatrix}\,,
\end{equation}
where we have also introduced the hybridization function
\begin{equation}\label{eq:gammafilter}
\Gamma_r(\omega)=\frac{\Sigma_r^A(\omega)-\Sigma_r^R(\omega)}{2}=-\mbox{Im}\Sigma_r^R(\omega)=\frac{t^2\Delta}{(\omega-\varepsilon_r)^2+\Delta^2}\,,
\end{equation}
which has the Lorentzian shape sketched in Figs. 1a and 2a of the main text.  

\subsection{Validity of the self-consistent Born scheme for the monitoring contribution to the self-energy }\label{sec:self_monitoring}

\begin{figure}
\begin{centering}
\includegraphics[width=\columnwidth]{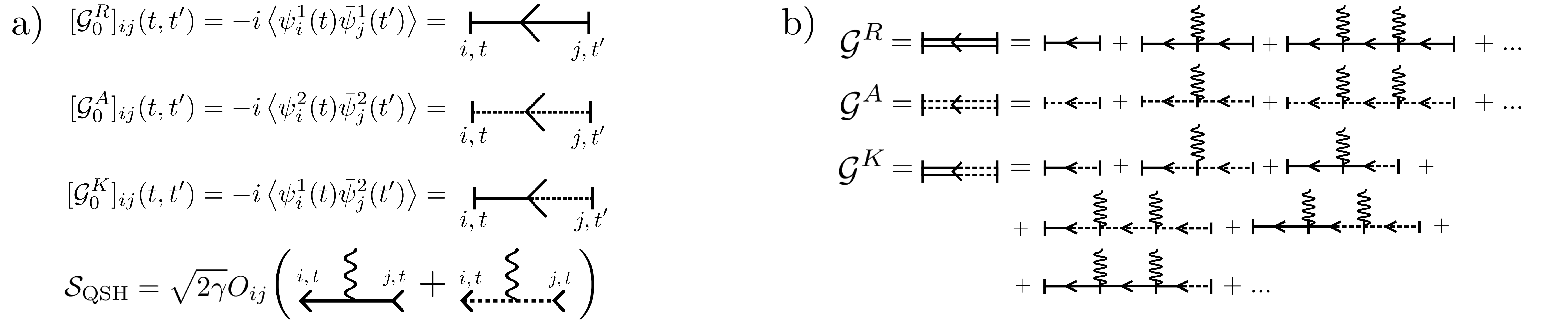}
\par\end{centering}
\caption{a)  Diagrammatic representation of the bare retarded, advanced and Keldysh Green functions and of the vertex associated to  the quantum stochastic action~\eqref{eq:stochasticaction}. Full lines represent the retarded propagator, dashed lines the advanced one, and mixed lines the Keldysh propagator. b)
\label{fig:proofexact}Diagrammatic expansion of the retarded, advanced and Keldysh propagators. At a given order $n$ in the expansion, only one diagram contributes to the retarded/advanced self-energy, whereas the Keldysh component has $n+1$ diagrammatic contributions, corresponding to the insertion of the Keldysh bare component at different times.}
\end{figure}

\begin{figure}
\begin{centering}
\includegraphics[width=\columnwidth]{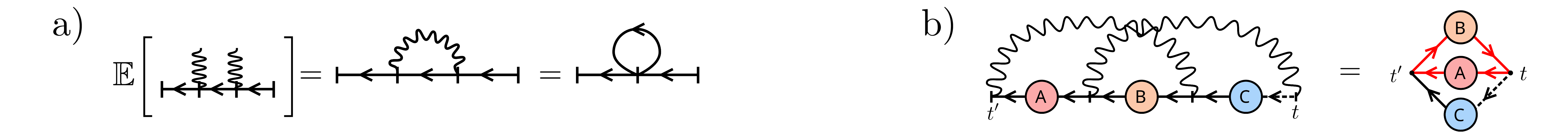}
\par\end{centering}
\caption{ \label{fig:proofexact2} a) Diagrammatic representation of the averaging over the noise realizations. The fact that the noise is Gaussian and $\delta$-correlated in time is represented by connecting two wiggly lines, which are then contracted to a single point in time.
b) Example of a crossing diagram for the Keldysh component. The red lines highlight the part
of the diagram violating the causality structure, as it features two retarded components with opposite directions in time. This diagram thus vanishes and can be discarded.
}
\end{figure}

The action~\eqref{eq:stoca}, associated to the monitoring of the observable $\mathcal O$ averaged over the measurement outcomes, is a quartic action in the Grassmann fields and thus cannot be integrated using Gaussian integrals. However, it is possible to derive an exact expression of the self-energy according to the self-consistent Born scheme, Eq.~(3) in the main text. The idea is to rely on the unraveling  procedure corresponding to Eq.~\eqref{eq:stochHamiltonian}. The corresponding action reads 
\begin{equation} \label{eq:stochasticaction}
\mathcal S_{{\rm QSH}} =  -\sum_{i,j}\int dt  \sqrt{2\gamma}O_{ij}\Big[\bar{d}_{i}^{1}(t)d_{j}^{1}(t)+\bar{d}_{i}^{2}(t)d_{j}^{2}(t)\Big] \xi(t) \,. 
\end{equation}
One can readily verify that performing the Gaussian noise average with moments $\mathbb E[\xi(t)]=0$ and $\mathbb E[\xi(t)\xi(t')]=\delta (t-t')$ directly leads to the dissipative action~\eqref{eq:stoca}
\begin{equation}
    \mathbb{E}[e^{i\mathcal S_{\rm sto}}]=\int D[\xi] e^{i\mathcal S_{\rm sto}-\int dt \xi^2(t)/2}=
    e^{i\mathcal S_{\cal D}}\,.
\end{equation}
The unraveling corresponding to Eq.~\eqref{eq:stochasticaction} thus corresponds to a Hubbard-Stratonovich transformation, where the  action becomes  quadratic in terms
of the Grassmann variables at the price of introducing the time-dependent noise $\xi(t)$. However, the fact that the {\it averaged} noise is $\delta$-correlated in time allows a dramatic simplification of  the diagrammatic expansion of the single-particle correlation functions in terms of 
the measurement strength $\gamma$.

Consider the diagrammatic representation of the Green's function in Fig.~\ref{fig:proofexact}a.  The vertex corresponding to Eq.~\eqref{eq:stochasticaction} only connects  solid lines with solid ones and dashed lines with dashed ones. As a consequence, in the diagrammatic expansion of the retarded (advanced) propagators only retarded (advanced) propagators appear. For the Keldysh component, one can switch only once from dashed to solid lines through the insertion of a Keldysh propagator, see Fig.~\ref{fig:proofexact}b. 

Averaging over noise realizations corresponds diagrammatically  to imposing an equal time index when connecting wiggly lines, see Fig.~\ref{fig:proofexact2}a. The key insight is that after noise-averaging the diagrams, those with crossing wiggly lines do not contribute to the action.  In Fig.~\ref{fig:proofexact2}b, we show an example of a diagram with crossing wiggly lines arising from the diagrammatic expansion of the Keldysh component. In that example, after averaging, two retarded propagators run in opposite time directions. As a consequence, this diagram involves the multiplication of two retarded functions with opposite time-dependence, which equals zero. Similar considerations apply for all crossing diagrams, and we redirect the interested reader to Refs.~\cite{Dolgirev_2020,Jin_2022} for the complete demonstration. Since the crossing diagrams vanish, the Born non-crossing approximation is exact, and all the remaining non-crossing diagrams can be exactly re-summed, leading to the  self-consistent equation for the self-energy
\begin{equation}\label{eq:sigma_all}
    \boldsymbol{\Sigma}^\gamma_{ij}(t,t')=2\g \d(t-t') \sum_{pq}O_{ip}\boldsymbol{\cal{G}}_{pq}(t,t) O_{qj}\,,
\end{equation}
which corresponds to Eq.~(3) in the main text.

\section{Derivation of the self-consistent Eq. (4) in the main text}

The way to solve the self-consistent equation~\eqref{eq:sigma_all} for its retarded and advanced components is given in the main text. We provide here additional details concerning the solution of its Keldysh component leading to Eq.~(4) in the main text.  We first recall that the total self-energy includes the contributions from the reservoirs. It has thus the form $\Sigma^K=\Sigma_R^K+\Sigma_L^K+2\gamma O\mathcal G^K\mathcal O$, with $\Sigma^K_{L/R}$ given in Eq.~\eqref{eq:selfres}. We then rely on the fact that $\mathcal G^K=\mathcal G^R\Sigma^K\mathcal G^A$ to derive, in time representation,
\begin{align}\label{eq:GKexact}
    \mathcal{G}^K(t,t)&=-2i\int\frac{d\w}{2\pi} \mathcal{G}^R(\omega) \left[\sum_{r=L/R}(1-2f_r(\omega))\G_r(\omega)\right]\mathcal{G}^A(\omega) +2\g\int\frac{d\w}{2\pi}\mathcal{G}^R(\omega)O\mathcal{G}^K(t,t) O\mathcal{G}^A(\omega)\;.
\end{align}
Relying on the identities 
\begin{align}\label{eq:ident}
\mathcal G^R-\mathcal G^A&=\mathcal G^R(\Sigma^R-\Sigma^A)\mathcal G^A=-2i\mathcal G^R\big(\Gamma_L+\Gamma_R+\gamma\mathcal O^2\big)\mathcal G^A\,,&\mathcal D_{ij}&=\langle d^\dagger_jd^{\phantom \dagger}_i\rangle=\frac{\d_{ij}-i \mathcal{G}^K_{ij}(t,t)}{2}\,,
\end{align} 
we can apply the property 
\begin{equation}
    i\int\frac{d\omega}{2\pi}[\mathcal G^R(\omega)-\mathcal G^A(\omega)]=\mathbb 1\,,
\end{equation} 
to derive a self-consistent equation for the correlation matrix $\mathcal D$, namely
\begin{equation}\label{eq:Dsapp}
\mathcal  D=\int\frac{d\w}{\pi} \mathcal{G}^R(\omega) \left[\sum_{r=L/R}f_r(\omega)\G_r(\omega)+\gamma O\mathcal DO\right]\mathcal{G}^A(\omega) \,,
\end{equation}
which corresponds to Eq. (4) in the main text. Equations of this type can be solved by vectorization or by direct substitution, for a few site systems. 

\section{Formula for currents in monitored systems -- Eq.~(5) of the main text}

In the presence of interactions or dissipation, the Landauer-B\"uttiker expressions~\cite{landauer_1957,Datta_1995} for elastic transport are no longer applicable and several extensions have been derived \cite{pastawski1992,Meir_1992}. We follow here the approach of Meir and Wingreen~\cite{Meir_1992,jin_generic_2020} and show how it can be used to derive Eq. (5) of the main text. 
We start by considering the particle ($\zeta=0$) and heat ($\zeta=1$) currents flowing into the reservoir $r$ -- Eq.~(2) in the main text
\begin{equation}\label{eq:currentsupmat}
J^{\zeta}_r=i\sum_{k,i} (\varepsilon_{r,k}-\mu_r)^{\zeta} \left[t_{r,ki}^{*} \langle d_i^{\dag} c^{\phantom \dagger}_{r,k} \rangle-t_{r,ki} \langle c^{\dagger}_{r,k} d^{\phantom \dagger}_i\rangle \right]=\sum_{k,i}\int\frac{d\omega}{4\pi}(\varepsilon_{r,k}-\mu_r)^{\zeta}\left[t_{r,ki}^{*} \mathcal G^K_{rk,i}(\omega)-t_{r,ki}\mathcal G^K_{i,rk}(\omega) \right]\,,
\end{equation}
where we have introduced the Keldysh correlation functions $\mathcal G^K_{i,kr}(t,t')=-i\langle[d^{\phantom\dagger}_i(t),c^\dagger_{r,k}(t')]\rangle$ and $\mathcal G^K_{kr,i}(t,t')=-i\langle[c^{\phantom\dagger}_{r,k}(t),d^\dagger_i(t')]\rangle$. By performing the diagrammatic expansion of these correlation functions in the tunnel amplitudes $t_{r,ki}$, one finds that they can be factorized as 
\begin{align}
    \mathcal G^K_{rk,i}(\omega)&=\mathcal C^R_{r,k}(\omega)\sum_j t_{r,kj}\mathcal G^K_{j,i}(\omega)+\mathcal C^K_{r,k}(\omega)\sum_jt_{r,kj}\mathcal G^A_{j,i}(\omega)\,,\label{eq:GKr1}\\
    \mathcal G^K_{i,rk}(\omega)&=\mathcal C^A_{r,k}(\omega)\sum_j t^*_{r,kj}\mathcal G^K_{i,j}(\omega)+\mathcal C^K_{r,k}(\omega)\sum_jt^*_{r,kj}\mathcal G^R_{i,j}(\omega)\,,\label{eq:GKr2}
\end{align}
where the correlation functions $\boldsymbol{\mathcal C}_{r,k}(\omega)$ are the correlation functions of the {\it isolated} reservoirs given in Eq.~\eqref{eq:CG0}, whereas the correlation functions $\boldsymbol{\mathcal G}$ are the full Green's functions of the system {\it including both reservoirs and monitoring}. By inserting Eqs.~(\ref{eq:GKr1}-\ref{eq:GKr2}) into Eq.~\eqref{eq:currentsupmat}, we then derive the expression
\begin{equation}
\begin{split}
J^\zeta_r&=\sum_{k,i,j}\int \frac{d\omega}{4\pi}(\varepsilon_{r,k}-\mu_r)^{\zeta}\,t_{r,ki}t^*_{r,kj}\Big[\Big(\mathcal C^R_{r,k}(\omega)-\mathcal C^A_{r,k}(\omega)\Big)\mathcal G^K_{i,j}(\omega)+\mathcal C^K_{r,k}(\omega)\Big(\mathcal G^A_{i,j}(\omega)-\mathcal G^R_{i,j}(\omega)\Big)\Big]\,.
\end{split}
\end{equation}
Exploiting the fact that $\mathcal C^R_{r,k}(\omega)-\mathcal C^A_{r,k}(\omega)=-2\pi i\delta(\omega-\varepsilon_{r,k})$ and that $\mathcal C^K_{r,k}(\omega)=-2\pi i \delta(\omega-\varepsilon_{r,k})\tanh[(\omega-\mu_r)/2T_r]$ we obtain:
\begin{equation}\label{eq:MW}
\begin{split}
J^\zeta_r&=-i\int \frac{d\omega}{2\pi}(\omega-\mu_r)^{\zeta}\,\mbox{tr}\Big[\Gamma_r(\omega)\mathcal G^K(\omega)+(1-2f_r(\omega))\Gamma_r(\omega)\Big(\mathcal G^A(\omega)-\mathcal G^R(\omega)\Big)\Big]\,,
\end{split}
\end{equation}
where we have introduced the generalized hybridization function $\Gamma_{r,ij}(\omega)=\pi\sum_k t^{\phantom *}_{r,ki}t^*_{r,kj}\delta(\omega-\varepsilon_{r,k})$. Note that although Eq.~\eqref{eq:gammafilter} corresponds to the particular case of a reservoir composed of a metallic lead tunnel coupled to an energy filter the relation $\Gamma_r(\omega)=[\Sigma^A_r(\omega)-\Sigma^R_r(\omega)]/2$ is always valid. 

The formula~\eqref{eq:MW} also applies to the monitored systems described by the Liouvillian dynamics~\eqref{eq:rhot}. Injecting the exact expressions Eqs.~(\ref{eq:GKexact}-\ref{eq:ident}), we  derive Eq.~(5) in the main text, namely
\begin{equation}\label{eq:mainsup}
J^{\zeta}_r=\frac{2}{\pi} \int d\w\,\ubcolor{ \left( \w -\mu_{r} \right)^{\zeta} \left( f_{\bar{r}} - f_{r} \right) \tr \left[ \G_r {\cal G}^R \G_{\bar{r}} {\cal G}^A \right] }{\text{elastic}} 
+\gamma\frac2\pi \int d\omega\,\ubcolor{\left( \w -\mu_{r} \right)^{\zeta} \tr \left[ \G_r {\cal G}^R O\left(\mathcal D-f_r \mathbb 1\right)O {\cal G}^A \right] }{\text{inelastic}}  \; .
\end{equation}
 
\section{Lindblad dynamics induced by a bosonic bath at large temperature and chemical potential}
 
In this section, we show how the Lindblad dynamics of Eq.~(1) in the main text can also describe the situation where a bosonic bath of large temperature and chemical potential is coupled to the monitored operator $\mathcal O$. This situation is described by a Hamiltonian of the form, $\mathcal H=\sum_k \varepsilon_k b_k^{\dag}b^{\phantom \dagger}_k+ \tau \sum_k (b^{\phantom \dagger}_k+b_k^{\dag})\mathcal O$, with the operator $\mathcal O$ being hermitian and quadatic $\mathcal O=\sum_{ij}d^\dagger_iO^{\phantom \dagger}_{ij}d^{\phantom \dagger}_j$, and where the operators $b^{\phantom \dagger}_k$ and $b^\dagger_k$ annihilate and create bosons of energy $\varepsilon_k$ in the bath. To derive the Keldysh action associated to this Hamiltonian, we follow the standard convention defining rotated classical and quantum bosonic fields in the Keldysh action~\cite{Kamenev_2011}  
\begin{align}
    b^{c}&=\frac{b^{+}+b^{-}}{\sqrt2}\,, &    b^{q}&=\frac{b^{+}-b^{-}}{\sqrt2}\,,
\end{align}
which also applies for the complex counterparts, at variance from the Grassman fermionic variables, see Eq.~\eqref{eq:grassman}. The resulting action reads
\begin{multline}\label{eq:actionboson}
     \mathcal S_{\rm boson~bath}=\sum_k\int dtdt' \left[\begin{array}{cc}
\bar b_{k}^{c} & \bar b_{k}^{q}\end{array}\right]_t [\boldsymbol{\mathcal{B}}_k(t-t')]^{-1}\left[\begin{array}{c}
b_{k}^{c}\\
b_{k}^{q}
\end{array}\right]_{t'}-\\ \tau\sum_{k,i,j}O_{i,j}\int dt\, \left[\frac{\bar b_k^c+b_k^c}{\sqrt2}(\bar d^1_id^1_j+\bar d^2_id^2_j)+\frac{\bar b_k^q+b_k^q}{\sqrt2}(\bar d^1_id^2_j+\bar d^2_id^1_j)\right]\,,
 \end{multline}
where $\boldsymbol{\mathcal{B}}_k$ is the Green's matrix of the isolated ($\tau=0$) bosonic bath, which in frequency space reads 
\begin{equation}
\boldsymbol{\mathcal B}_k(\omega)=\begin{bmatrix}
\mathcal B_k^K(\omega)&\mathcal B_k^R(\omega)\\
\mathcal B_k^A(\omega)&0
\end{bmatrix}
=
\begin{bmatrix}
-2i\pi\delta(\omega-\varepsilon_k)\coth\left(\frac{\omega-\mu_B}{2T_B}\right)&\frac{1}{\omega-\varepsilon_k+i0^+}\\
\frac{1}{\omega-\varepsilon_k-i0^+}&0
\end{bmatrix}\,,
\end{equation}
where $\mu_B$ and $T_B$ are the chemical potential and the temperature of the bosonic bath respectively. 
The bosonic degrees of freedom of the action~\eqref{eq:actionboson} can be integrated using a Gaussian integral to obtain
\begin{equation}\label{eq:thermal}
\begin{split}
\mathcal S'_{\text{boson bath}}	&=-\frac{\tau^2}{2}\sum_{k,i,j,k,l}O_{ij}O_{kl}\int dt dt' \left[\begin{array}{cc}
\bar d_i^1d_j^1+\bar d_i^2 d_j^2 & \bar d_i^1d_j^2+\bar d_i^2 d_j^1\end{array}\right]_t
\boldsymbol{\mathcal B}_k(t-t')\left[\begin{array}{c}\bar d_k^1d_l^1+\bar d_k^2 d_l^2\\
\bar d_k^1d_l^2+\bar d_k^2 d_l^1 \end{array}\right]_{t'}\\
&=-\frac{\tau^2}{2}\sum_{k,i,j,k,l}O_{ij}O_{kl}\int dt dt'\Big\{ 
\Big(\bar d_i^1d_j^1+\bar d_i^2 d_j^2\Big)_t\mathcal B_k^K(t-t')\Big(\bar d_k^1d_l^1+\bar d_k^2 d_l^2\Big)_{t'}+ \\
&~~~~~~~~~~~~~~~~~~~~~~~~~~~~~~~~~~~~~~~~~~~~ \Big(\bar d_i^1d_j^1+\bar d_i^2 d_j^2\Big)_t\Big[\mathcal B^R_k(t-t')+\mathcal B^A_k(t'-t)\Big]\Big(\bar d_k^1d_l^2+\bar d_k^2 d_l^1\Big)_{t'}\Big\}\,.
\end{split}
\end{equation}
We want to find the conditions for which this action maps onto the action~\eqref{eq:stoca}, corresponding to the Lindbladian in Eq.~\eqref{eq:rhot}. For this, the last term in Eq.~\eqref{eq:thermal} must vanish and  $\sum_k\mathcal B^K_k(t-t')$ is proportional to a $\delta$-function in time, or, equivalently, its Fourier transform does not depend on frequency. The last term in Eq.~\eqref{eq:thermal} is proportional to 
\begin {equation}\label{eq:RA}
\begin{split}
\sum_k\Big[\mathcal B^R_k(t-t')+\mathcal B^A_k(t'-t)\Big]&=-i\theta(t-t')\sum_k\Big[e^{-i\varepsilon_k(t-t')}-e^{i\varepsilon_k(t-t')}\Big]\\&
=-i\theta(t-t')\int d\omega\,\nu(\omega)\Big[e^{-i\omega(t-t')}-e^{i\omega(t-t')}]\,,
\end{split}
\end{equation}
where we have introduced the density of states of the bosonic bath $\nu(\omega)$. If this density of states is constant, the contribution~\eqref{eq:RA} vanishes as well as the last term in Eq.~\eqref{eq:thermal}.

The $k$-integration of the Keldysh component gives instead 
\begin{equation}
\sum_k\mathcal B^K(t)=-i \int d\w\, \nu(\w) \coth \left( \frac{\w-\mu_B}{2T_B}\right) e^{-i \w t}\,,
\end{equation}
which would equal $\sum_k\mathcal B^K(t)\simeq 2\pi i\coth(\mu_B/2T_B)\delta(t)$, if we can neglect the energy dependence of the density of states $\nu(\omega)$ and of the cotangent in the integration. Assuming negative chemical potential, one finds the action~\eqref{eq:stoca} by making the identification 
\begin{equation}
\gamma=\pi \tau^2\coth\left(\frac{|\mu_B|}{2T_B}\right)\,.
\end{equation}
The above approximations can be justified in the limit where the density of states of the  bosonic bath is constant for energies comparable to the spectral width of the fermionic system to which the bosonic bath is coupled to, and in the limit where $\mu_B$ and $T_B$ are much larger than the spectral width of the fermionic system. Similar arguments were used to derive loss and gain terms in the Lindblad form by allowing the system to exchange particles with a fermionic reservoir with a much larger temperature and chemical potential~\cite{jin_generic_2020}.

\end{document}